# Statistical Fluid Mechanics: Dynamics Equations and Linear Response Theory

Haibing Peng*

NOR-MEM Microelectronics Co., Ltd., Suzhou Industrial Park, Suzhou, JiangSu Province 215000, P.R. China

*Permanent Email: haibingpeng@post.harvard.edu

ABSTRACT

The statistical nature of discrete fluid molecules with random thermal motion so far has not been considered in mainstream fluid mechanics based on Navier-Stokes equations, wherein fluids have been treated as a continuum breaking into many *macroscopically* infinitely small (but *microscopically* large enough) mass elements with their motion only characterized by center-of-mass velocity. Here we provide a Statistical Mechanical approach to address fluid dynamics by considering statistical velocity distribution of discrete molecules within *macroscopically* infinitely small volume elements as well as their center-of-mass velocity. Dynamics equations governing the evolution of physical variables have been proposed, Green's functions have been obtained and linear response theory has been applied to study physical situations with external heat perturbation. It is found that the propagation of heat, center-of-mass motion and sound are intrinsically integrated in Statistical fluid dynamics. This work lays down the foundation for applications of Statistical fluid mechanics.

It is well understood from Statistical Mechanics that fluids consist of an assembly of discrete individual molecules with random thermal motion, *e.g.*, in gaseous fluids such as air, molecules move on average at about the speed of sound $V_s$ ~340 *m/s* at room temperature. However, the statistical nature of discrete fluid molecules with random thermal motion so far has not been incorporated into mainstream fluid mechanics based on Navier-Stokes (NS) equations,[1, 2] wherein the fluid dynamics has been governed by Newton's second law, and a fluid has been treated as a continuum breaking into many *macroscopically* infinitely small, but *microscopically* so large (on the scale of molecule size) mass elements with their motion only characterized by center-of-mass velocity. In a previous work,[3] we introduced a Statistical Mechanical approach to study the air-solid interface and resultant aerodynamic forces. Here we apply principles of Statistical Mechanics to address the dynamics inside fluids, by considering statistical velocity distribution of discrete molecules within *macroscopically* infinitely small but *statistically* large enough volume elements as well as their center-of-mass velocity. Equations of statistical fluid dynamics have been proposed and linear response theory has been applied to obtain general solutions for such dynamic equations.

We start by breaking a fluid into many small volume elements in position space $d\Omega \equiv d^3\boldsymbol{r}$, which can be treated as *macroscopically* infinitely small but *microscopically* large enough to contain many discrete molecules accountable by statistical distributions. Every volume element is characterized by a time-dependent molecule number density in position and velocity phase space:

$$\psi(t,\boldsymbol{r},\boldsymbol{V}) \equiv n(\boldsymbol{r},t)\rho(\boldsymbol{V},\boldsymbol{r},t), \qquad (1)$$

where $n(\boldsymbol{r},t)$ is the molecule number density in position space, and $\rho(\boldsymbol{V},\boldsymbol{r},t)$ is proportional to the probability density in velocity space which we choose to be an unnormalized dimensionless form, *e.g.*, $\rho(\boldsymbol{V},\boldsymbol{r},t) = e^{-mV^2/(2k_BT)}$ for Boltzmann distribution at temperature T, with $m$ being the molecular mass and $k_B$ the Boltzmann constant. For any classical physical variable (without considering spin) $f = f(\boldsymbol{r},\boldsymbol{V})$, its average value at time $t$ can be obtained as:

$$<f(t)> = \frac{\int_\Omega d^3r \int_{\Omega_V} d^3V \; f(\boldsymbol{r},\boldsymbol{V}) \cdot \psi(t,\boldsymbol{r},\boldsymbol{V})}{\int_\Omega d^3r \int_{\Omega_V} d^3V \; \psi(t,\boldsymbol{r},\boldsymbol{V})}, \qquad (2)$$

wherein the integration is taken over all applicable volume $\Omega$ in position space and $\Omega_V$ in velocity space. Therefore, the dynamics of $\psi(t,\boldsymbol{r},\boldsymbol{V})$ carries important information of the fluids which can be used to trace the dynamics of any physical variables.

In general, depending on elastic properties of fluids and the strength of external perturbations, fluids can be classified into two categories: incompressible and compressible. For incompressible fluids, $n(\boldsymbol{r},t) = n(\boldsymbol{r})$ is time-independent; we can treat every volume element as an *effective* canonical ensemble since despite molecules moving in or out the overall physical result is that the molecule number inside the volume element is unchanged but its summed energy fluctuates on an average energy with the rest of the whole fluid system being treated as a reservoir at fixed temperature *T*, which is effectively the same as the typical canonical ensemble [4] where there is no exchange of molecules but exists exchange of energy with a reservoir at fixed *T*. Therefore, in incompressible regime, $\rho(\boldsymbol{V},\boldsymbol{r},t)$ is characterized by an effective local temperature $T_e(\boldsymbol{r},t)$. For compressible fluids, $n(\boldsymbol{r},t)$ is time-dependent; we can treat every volume element as a grand canonical ensemble [4] with exchange of molecules and its energy fluctuate on an average energy with the rest of the whole fluid system being treated as a reservoir at fixed *T*; therefore, in compressible regime, $\rho(\boldsymbol{V},\boldsymbol{r},t)$ is characterized by an effective local temperature $T_e(\boldsymbol{r},t)$ and a local chemical potential $\mu_e(\boldsymbol{r},t)$.

Next we address the dynamical evolution of $\psi(t,\boldsymbol{r},\boldsymbol{V})$ by examining its flux density $\boldsymbol{J}(t,\boldsymbol{r},\boldsymbol{V})$ associated with the volume element in position space $d\Omega$. Since the physical meaning of $\psi(t,\boldsymbol{r},\boldsymbol{V})$ is the number density of a group of molecules with the same velocity $\boldsymbol{V}$, in the case with no external perturbation to provide any source of creation/annihilation of molecules inside a fluid, we immediately have the dynamics equation as the continuity equation:

$$\frac{\partial \psi(t,\boldsymbol{r},\boldsymbol{V})}{\partial t} = -\boldsymbol{\nabla} \cdot \boldsymbol{J}(t,\boldsymbol{r},\boldsymbol{V}), \qquad (3)$$

where the gradient operator acts on position variable $\boldsymbol{r}$.

In *ballistic regime* where intermolecular scattering is negligible (*e.g.*, at low temperature limit or for certain special fluid molecules with extremely small scattering cross-section so that the length scale of interest for study is far less than the mean free path $l_m$ for intermolecular scattering), the flux density is given by

$$\boldsymbol{J}(t,\boldsymbol{r},\boldsymbol{V}) = \boldsymbol{V}\psi(t,\boldsymbol{r},\boldsymbol{V}), \qquad (4)$$

and thus we obtain the dynamics equation for ballistic regime as:

$$\frac{\partial \psi(t,r,V)}{\partial t} + V \cdot \nabla \psi(t,r,V) = 0 . \tag{5}$$

In *diffusive regime*, intermolecular scattering is significant in fluids, *e.g.*, when the length scale of interest for study is far larger than the mean free path $l_m$. Consequently, although the actual instantaneous flux density is still given by eq. (*4*), the target molecules relevant to $\psi(t,r,V)$ physically experience random walk (Brownian motion) due to intermolecular scattering, and thus the effective flux density associated with the volume element $d\Omega$ should be given by diffusion theory as:

$$J(t,r,V) = -D(V)\nabla \psi(t,r,V). \tag{6}$$

Here the diffusion constant is dependent on the speed $V$ and given by $D(V) = V l_m/2$ according to Einstein-Smoluchowski relation,[5, 6] where the random walk step distance has been taken as the mean free path $l_m$ for intermolecular scattering. If the scattering probability per unit volume $p$ is uniform all over the space, the mean free path $l_m$ should be independent of the speed $V$ of random-walking target molecules relevant to $\psi(t,r,V)$ because the scattering probability for target molecules to travel a distance of $l_m$ is $P = p \cdot \delta \cdot l_m$ with $\delta$ the cross sectional scattering area for the molecules. In this work, we only consider the case for classical fluids with positive real diffusion constants; the cases for general fluids with time-dependent diffusion flux density $J(t,r,V)$ having a phase difference with $\psi(t,r,V)$ are discussed elsewhere.[7, 8]

From eq. (*3*) and (*6*),[9] we then obtain the dynamics equation for diffusive regime as:

$$\frac{\partial \psi(t,r,V)}{\partial t} - D(V)\nabla^2 \psi(t,r,V) = 0 . \tag{7}$$

The trivial steady state solution of eq. (*7*) is a uniform state $\psi_0$, e.g., a thermal equilibrium state at constant temperature $T_0$ satisfying Boltzmann distribution; and its eigenstates are non-steady damping waves:

$$\psi_k(t,r,V) = e^{-\alpha(k)t + ik \cdot r}, \tag{8}$$

with a dispersion relation $\alpha(k) = D(V)k^2$. Mathematically, the general solution to eq. (*7*) can be expressed as:

$$\psi(t,r,V) = \psi_0 + \int c(k)\psi_k(t,r,V)d^3k, \tag{9}$$

where the coefficients $c(k)$ are determined by initial and boundary conditions and should also satisfy $c(-k) = c^*(k)$ in order to guarantee that $\psi(t,r,V)$ is a real function.

However, for most classical fluids, the *normal regime* is usually between the ballistic and the diffusive limits, where the effective flux density should be a combination of the drift flux density of eq. (4) and the diffusion flux density of eq. (6):

$$J(t,r,V) = V\psi(t,r,V) - D(V)\nabla \psi(t,r,V). \tag{10}$$

We note that in the normal regime, the effective drift flux density may be reduced by scattering to $\xi V\psi(t,r,V)$ with an attenuation factor $0 < \xi \leq 1$, and we take $\xi = 1$ in eq. (*10*). With that, the dynamics equation for the normal regime is:

$$\frac{\partial \psi(t,r,V)}{\partial t} + V \cdot \nabla \psi(t,r,V) - D(V)\nabla^2 \psi(t,r,V) = 0 . \tag{11}$$

Its eigenstates can still be written as eq. (8) with a dispersion relation $\alpha(k) = D(V)k^2 + iV \cdot k$, which are now non-steady damping plane waves with oscillations in time.

In order to gain better physical insight, below we set up practical physical situations and apply linear response theory to obtain solutions. Suppose we have a fluid system (*e.g.*, a gaseous atmospheric system) initially in a steady state

$$\psi_0 = n_0 e^{-mV^2/(2k_B T_0)} \equiv n_0 \rho_0, \tag{12}$$

with $n_0$ the molecule number density in position space. This initial state is uniform in position space and in thermal equilibrium at temperature $T_0$ characterized by Boltzmann distribution. At certain time, an external perturbation $\Delta\psi(t,\boldsymbol{r},\boldsymbol{V})$ is turned on to allow energy and/or molecules flow into or out of the fluid system locally at certain positions, and response of the fluid system is reflected by the dynamic evolution of $\psi(t,\boldsymbol{r},\boldsymbol{V})$. In order to determine which regime is suitable to use, the length scale of interest for study is the distance from the perturbation position to the target position $\boldsymbol{r}$: the diffusive regime is far-field compared with the mean free path while the ballistic regime is near-field. Also, since the external perturbation $\Delta\psi(t,\boldsymbol{r},\boldsymbol{V})$ introduces an extra source of creation/annihilation of molecules inside the fluid, the continuity equation should be modified by adding the resultant molecule creation rate $\partial\{\Delta\psi(t,\boldsymbol{r},\boldsymbol{V})\}/\partial t$ to the right hand side (R.H.S.) of eq. (3). Therefore, with external perturbation $\Delta\psi(t,\boldsymbol{r},\boldsymbol{V})$ as a driving force, the dynamics equation should be modified from eq. (*5*), (*7*) and (*11*) for applicable regimes, respectively, by adding a driving term $\partial\{\Delta\psi(t,\boldsymbol{r},\boldsymbol{V})\}/\partial t$ to the R.H.S..

In diffusive regime, the dynamics equation under driving is:
$$\frac{\partial\psi(t,\boldsymbol{r},\boldsymbol{V})}{\partial t} - D(\boldsymbol{V})\boldsymbol{\nabla}^2\psi(t,\boldsymbol{r},\boldsymbol{V}) = \frac{\partial\{\Delta\psi(t,\boldsymbol{r},\boldsymbol{V})\}}{\partial t}. \qquad (13)$$
For a fluid system infinitely large in size (*i.e.*, without a boundary), the Fourier transform of the Green's function,[10] defined as the response under a driving term $\delta(t)\delta^3(\boldsymbol{r})$, can be obtained immediately by applying this driving term to the R.H.S. of eq. (13) and performing Fourier transform over the time and position spaces:
$$G(\omega,\boldsymbol{k},\boldsymbol{V}) = \frac{1}{D(\boldsymbol{V})k^2 - i\omega}, \qquad (14)$$
with $\omega$ being the frequency and $\boldsymbol{k}$ the wave vector. The Green's function $g(t,\boldsymbol{r},\boldsymbol{V})$ in Gaussian form can be given following standard diffusion theory [1] as:
$$g(t,\boldsymbol{r},\boldsymbol{V}) = \{4\pi D(\boldsymbol{V})t\}^{-3/2} \cdot e^{-\frac{r^2}{4D(\boldsymbol{V})t}} \cdot \theta(t), \qquad (15)$$
where $\theta(t)$ is the step function. For any arbitrary driving function $f(t,\boldsymbol{r})$, the response is the convolution between it and the Green function of eq. (15):
$$\psi(t,\boldsymbol{r},\boldsymbol{V}) = \psi_0 + f(t,\boldsymbol{r}) * g(t,\boldsymbol{r},\boldsymbol{V})$$
$$= \psi_0 + \int_{-\infty}^{t} d\tau \int_{\Omega} d^3\boldsymbol{r}' f(\tau,\boldsymbol{r}')g(t-\tau,\boldsymbol{r}-\boldsymbol{r}',\boldsymbol{V}). \qquad (16)$$

However, numerical evaluation of the integration in eq. (16) is difficult if $f(t,\boldsymbol{r})$ is oscillating at a set of discrete frequencies. Therefore, we provide below a convenient method for solving eq. (13) under the driving term $f(t,\boldsymbol{r}) = e^{-i\omega_0 t} f_1(\boldsymbol{r})$ with $f_1(\boldsymbol{r})$ an arbitrary function. First, by applying a driving term $e^{-i\omega_0 t}\delta^3(\boldsymbol{r})$ to the R.H.S. of eq. (13) and performing Fourier transform, we have
$$-i\omega\,\Psi(\omega,\boldsymbol{k},\boldsymbol{V}) + D(\boldsymbol{V})k^2\Psi(\omega,\boldsymbol{k},\boldsymbol{V}) = \delta(\omega-\omega_0),$$
wherein $\Psi$ is the Fourier transform of the response $\psi$. This leads to
$$\Psi(\omega,\boldsymbol{k},\boldsymbol{V}) = \frac{1}{D(\boldsymbol{V})k^2 - i\omega_0}\delta(\omega-\omega_0). \qquad (17)$$
After that, inverse Fourier transform of eq. (*17*) is performed (with integration in $\boldsymbol{k}$ space done using spherical coordinates and further with mathematical method of contour integration), and then we obtain the solution for the response as:
$$\psi(t,\boldsymbol{r},\boldsymbol{V}) = \psi_0 + \frac{1}{D(\boldsymbol{V})}\frac{e^{-k_0 r}e^{ik_0 r}}{4\pi r}e^{-i\omega_0 t}. \qquad (18)$$
Here the solution is a spatially damping concentric radial wave (note that it is like the screened electrostatic Coulomb potential but modified with an oscillation along radial direction), and the damping wave vector is

$$k_0 \equiv \sqrt{\frac{\omega_0}{2D(V)}} = \sqrt{\frac{\omega_0 \tau}{l_m^2}}, \tag{19}$$

with the molecular relaxation time $\tau = l_m/V$ and $D(V) = Vl_m/2$. By defining a kernel function:

$$w(\boldsymbol{r}, \boldsymbol{V}) = \frac{1}{D(V) \cdot 4\pi r} e^{-k_0 r} e^{ik_0 r}, \tag{20}$$

we then readily obtain the response to the driving $f(t, \boldsymbol{r}) = e^{-i\omega_0 t} f_1(\boldsymbol{r})$ as

$$\psi(t, \boldsymbol{r}, \boldsymbol{V}) = \psi_0 + \{f_1(\boldsymbol{r}) * w(\boldsymbol{r}, \boldsymbol{V})\} e^{-i\omega_0 t}, \tag{21}$$

where the convolution is performed in $\boldsymbol{r}$ space.

Next, we look at the normal regime, where the dynamics equation under driving is:

$$\frac{\partial \psi(t, \boldsymbol{r}, \boldsymbol{V})}{\partial t} + \boldsymbol{V} \cdot \nabla \psi(t, \boldsymbol{r}, \boldsymbol{V}) - D(V) \nabla^2 \psi(t, \boldsymbol{r}, \boldsymbol{V}) = \frac{\partial \{\Delta \psi(t, \boldsymbol{r}, \boldsymbol{V})\}}{\partial t}. \tag{22}$$

The Fourier transform of the Green's function for an infinitely large fluid is:

$$G(\omega, \boldsymbol{k}, \boldsymbol{V}) = \frac{1}{D(V)k^2 + i\boldsymbol{V} \cdot \boldsymbol{k} - i\omega}. \tag{23}$$

We then rewrite eq. (23) as:

$$G(\omega, \boldsymbol{k}, \boldsymbol{V}) = \frac{1}{D(V)|\boldsymbol{k} + i\frac{\boldsymbol{V}}{2D(V)}|^2 - i\{\omega + i\left(\frac{V^2}{4D(V)}\right)\}},$$

Comparing with eq. (14) and (15), we can immediately obtain the Green's function here by multiplying the result of eq. (15) by a factor $e^{\frac{\boldsymbol{V} \cdot \boldsymbol{r}}{2D(V)}} \cdot e^{-\frac{V^2}{4D(V)}t}$ using the translational theorem for Fourier transform, resulting in:

$$g(t, \boldsymbol{r}, \boldsymbol{V}) = \{4\pi D(V)t\}^{-3/2} \cdot e^{-\frac{r^2}{4D(V)t} + \frac{\boldsymbol{V} \cdot \boldsymbol{r}}{2D(V)} - \frac{V^2}{4D(V)}t} \cdot \theta(t)$$
$$= \{4\pi D(V)t\}^{-3/2} \cdot e^{-\frac{|\boldsymbol{r} - \boldsymbol{V}t|^2}{4D(V)t}} \cdot \theta(t). \tag{24}$$

This is an intriguing but physically reasonable result as it describes a diffusion Gaussian wave packet with its center moving at velocity $\boldsymbol{V}$. Similarly, under a driving term $e^{-i\omega_0 t} \delta^3(\boldsymbol{r})$, we obtain the response kernel function as

$$w(\boldsymbol{r}, \boldsymbol{V}) = \frac{1}{D(V) \cdot 4\pi r} e^{-k_1 r} e^{ik_1 r} e^{\frac{\boldsymbol{V} \cdot \boldsymbol{r}}{2D(V)}}, \tag{25}$$

with the complex wave vector defined as

$$k_1 \equiv \sqrt{\frac{\omega_0 + i\left(\frac{V^2}{4D(V)}\right)}{2D(V)}}. \tag{26}$$

Now we are ready to look at practical physical examples, with the Green's functions and the response kernel functions derived above for both the diffusive and the normal regimes. Typical external perturbations include thermal heat transfer, external mechanical forces, and exchange of molecules via diffusion or chemical reactions. Below we will mainly deal with the response of incompressible fluids under the first two types of external perturbations mentioned above (we will address the response of incompressible fluids in a separate study).

We assume that the external perturbation is not strong enough so that the fluid system can be treated as incompressible (*e.g.*, ambient atmospheric system should be incompressible with an external pressure perturbation less than $10^5$ Pa). In such cases, the perturbation-induced response is usually a non-steady decaying state

$$\psi(t, \mathbf{r}, \mathbf{V}) = n_0 \rho(\mathbf{V}, \mathbf{r}, t), \tag{27}$$

where we propose that $\rho(\mathbf{V}, \mathbf{r}, t)$ can be *approximately* described by a local quasi-equilibrium state (Fig. 1)

$$\rho(\mathbf{V}, \mathbf{V}_e, T_e) = e^{-m|\mathbf{V} - \mathbf{V}_e|^2 / (2k_B T_e)}, \tag{28}$$

with an effective temperature $T_e(\mathbf{r}, t)$ and an effective center-of-mass displacement velocity $\mathbf{V}_e(\mathbf{r}, t)$. More explicitly, with the principle of eq. (2), these two parameters can be numerically extracted from:

$$\mathbf{V}_e(\mathbf{r}, t) \cong \frac{\int_{\Omega_V} \mathbf{V} \rho(\mathbf{V}, \mathbf{r}, t) \, d^3 V}{\int_{\Omega_V} \rho(\mathbf{V}, \mathbf{r}, t) \, d^3 V}, \tag{29}$$

and

$$Z = \int_{\Omega_V} \rho(\mathbf{V}, \mathbf{r}, t) \, d^3 V \cong (2\pi k_B T_e / m)^{3/2}, \tag{30}$$

where $Z$ is the partition function.

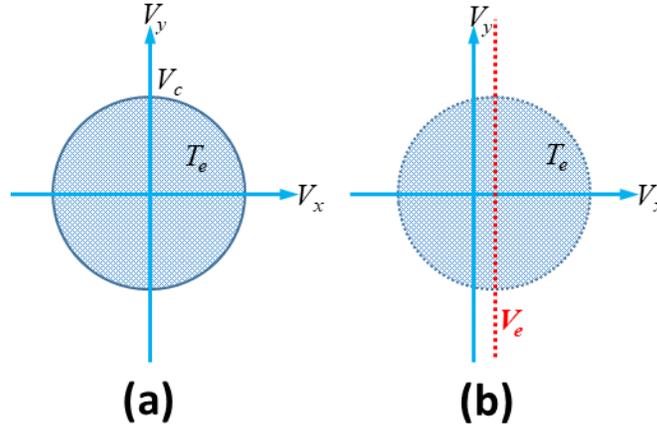

**Fig. 1.** Statistical distribution of molecules in velocity space filled up to a cutoff speed $V_c$ for: **(a)** an equilibrium state and **(b)** a quasi-equilibrium state, with $T_e$ the effective temperature and $\mathbf{V}_e$ the effective center-of-mass displacement velocity.

As the first physical example to study, we choose a fluid system infinitely large in size (*i.e.*, without a boundary) and apply time-dependent local heat flow as the perturbation so that the local effective temperature is driven at a monotonic frequency $\omega_0$:

$$\Delta T_e(\mathbf{r}, t) = \gamma e^{-i\omega_0 t} \delta^3(\mathbf{r}). \tag{31}$$

Note that by convention we take the real part of complex functions to be the value of physical variables. From eq. (27) and (28) we have

$$\Delta \psi(t, \mathbf{r}, \mathbf{V}) = \Delta T_e(\mathbf{r}, t) \cdot \frac{\partial \psi(t, \mathbf{r}, \mathbf{V})}{\partial T_e} \Big|_{T_e = T_0} = \Delta T_e(\mathbf{r}, t) \cdot \psi_0 \cdot mV^2 / (2k_B T_0^2), \tag{32}$$

with the initial steady state $\psi_0$ given by eq. (*12*).

To investigate far-field response, we use the dynamics equation under driving for diffusive regime, eq. (13), and by applying the driving-force term of eq. (*32*) we then have

$$\frac{\partial \psi(t, \mathbf{r}, \mathbf{V})}{\partial t} - D(V) \nabla^2 \psi(t, \mathbf{r}, \mathbf{V}) = AV^2(-i\omega_0) e^{-i\omega_0 t} \delta^3(\mathbf{r}), \tag{33}$$

where the constant $A \equiv \psi_0 \cdot m\gamma / (2k_B T_0^2)$. From eq. (*18*) or (*21*), with the initial steady state $\psi_0$ and $D(V) = V l_m / 2$, we can easily reach a solution:

$$\psi(t, \boldsymbol{r}, \boldsymbol{V}) = \psi_0 + \frac{AV(-i\omega_0)}{2\pi l_m} \frac{e^{-k_0 r} e^{ik_0 r}}{r} e^{-i\omega_0 t}, \tag{34}$$

where the damping wave vector $k_0$ is given in eq. (19). As can be seen from eq. (19) and (34), if the driving frequency $\omega_0$ is near $1/\tau$ (with $\tau$ being the molecular relaxation time), the fluid response is exponentially damped radially in a distance $\sim l_m$, yet at low frequency limit $\omega_0 \ll 1/\tau$ the response can extend much further away from the center.

Now we are ready to compute the response of $T_e$ and $\boldsymbol{V}_e$. Based on eq. (32) and (34), we can easily obtain the self-response of $T_e$ as:

$$\begin{aligned} T_e(\boldsymbol{r}, t) &= T_0 + \frac{AV(-i\omega_0)}{2\pi l_m} \frac{e^{-k_0 r} e^{ik_0 r}}{r} e^{-i\omega_0 t} \cdot \frac{2 k_B T_0^2}{\psi_0 \cdot mV^2} \\ &= T_0 + \gamma \frac{(-i\omega_0)}{4\pi D(V)} \frac{e^{-k_0 r} e^{ik_0 r}}{r} e^{-i\omega_0 t} \ . \end{aligned} \tag{35}$$

The result indicates that damping heat waves are generated which presents an intriguing spatial oscillation along radial direction. In addition, sound waves are also produced as the local pressure depends on the temperature, which is interesting since although the fluid is incompressible with a constant $n_0$, the molecular number density is redistributed in velocity space. We note that the dispersion relation for both the heat wave and the sound wave is the same and shown in eq. (19) as, $\omega_0 = 2D(V) k_0^2$, which could be explored in experiments to determine the diffusion constant.

The response of $\boldsymbol{V}_e$ can be determined by using the results of eq. (34) and (27) for the integration of eq. (29). Note that from eq. (34), we have $\{\psi(t, \boldsymbol{r}, \boldsymbol{V}) - \psi_0\} = n_0\{\rho(\boldsymbol{V}, \boldsymbol{r}, t) - \rho_0\}$ proportional to $V\rho_0$, therefore from eq. (29) we can deduce that the linear response term

$$\boldsymbol{V}_e(\boldsymbol{r}, t) \propto \int_{\Omega_V} \boldsymbol{V} V e^{-mV^2/(2k_B T_0)} \ d^3 \boldsymbol{V} = 0 \ ,$$

considering the symmetry of the integral in $\boldsymbol{V}$ space. This indicates that there is no center-of-mass motion in the far-field region with an perturbation oscillating in time locally, *i.e.*, $e^{-i\omega_0 t} \delta^3(\boldsymbol{r})$.

Next, we examine the response of the same perturbation of eq. (*31*) in the normal regime. With the response kernel function given in eq. (25), we immediately obtain

$$T_e(\boldsymbol{r}, t) = T_0 + \gamma \frac{(-i\omega_0)}{4\pi D(V)} \frac{e^{-k_1 r} e^{ik_1 r} e^{\frac{\boldsymbol{V} \cdot \boldsymbol{r}}{2D(V)}}}{r} e^{-i\omega_0 t} \ . \tag{36}$$

Compared with the far-field result of eq. (35) which is isotropic in heat propagation direction and only dependent on radial distance $r$, the response of $T_e(\boldsymbol{r}, t)$ in the normal regime now shows anisotropic heat propagation preferred along the direction of $\boldsymbol{V}$, which is shown by the term $e^{\frac{\boldsymbol{V} \cdot \boldsymbol{r}}{2D(V)}}$ in eq. (36). In addition, the response of $\boldsymbol{V}_e$ is now in a form

$$\boldsymbol{V}_e(\boldsymbol{r}, t) \propto \int_{\Omega_V} \boldsymbol{V} V e^{\frac{\boldsymbol{V} \cdot \boldsymbol{r}}{2D(V)}} e^{-mV^2/(2k_B T_0)} \ d^3 \boldsymbol{V}. \tag{37}$$

By Taylor expansion of the $e^{\frac{\boldsymbol{V} \cdot \boldsymbol{r}}{2D(V)}}$ term in the integral, we find that the symmetry of the integral in $\boldsymbol{V}$ space is broken, and $\boldsymbol{V}_e(\boldsymbol{r}, t)$ is nonzero. Therefore, macroscopic center-of-mass fluid motion (*i.e.*, convection) is indeed generated by the oscillating heat perturbation in the normal regime, although such a thermal-mechanical motion disappears in the far field (diffusive regime). This provides an integrated physical picture that heat transfer in fluids is carried out via both convection and diffusion in near field (normal regime), but dominantly via diffusion in the far field (diffusive regime).

To summarize, in this work we have constructed the framework of Statistical fluid mechanics [11-14] by providing the dynamics equations governing the evolution of $\psi(t, \boldsymbol{r}, \boldsymbol{V})$, the molecule number density in position and velocity phase space, which is of paramount importance for studying evolution of all physical variables. Furthermore, we have given the Green's functions for solving the dynamics equations and applied the linear response theory in the study of practical physical situations with external heat perturbation. It is found that the propagation of heat, center-of-mass motion and sound are intrinsically integrated in Statistical fluid dynamics. This work has laid down the foundation for applications of Statistical mechanics in fluid systems. In addition, the method can be generated to Boson and Fermi gases, *e.g.*, to study electron transport and thermoelectric effect for electron gases in solids.

## Acknowledgements

We would like to thank X. Peng for discussions on Oriental Philosophy of Zen which has stimulated the idea of dynamics in this work.